\documentclass[runningheads]{llncs}
\usepackage[T1]{fontenc}
\usepackage{graphicx}
\usepackage{framed}

\usepackage[utf8]{inputenc}
\usepackage[english]{babel}
\usepackage{setspace}
\usepackage{graphicx}
\usepackage{amsmath,amssymb,amsfonts}
\usepackage{algorithmic}
\usepackage{graphicx}
\usepackage{textcomp}
\usepackage{xcolor}
\usepackage{xspace}
\usepackage{verbatim}
\usepackage{wrapfig}
\usepackage{subfigure}
\usepackage{hyperref}
\usepackage{todonotes}
\usepackage{mathptmx}
\usepackage{amssymb}
\usepackage{placeins}
\usepackage{url}
\usepackage{multirow}
\usepackage{xcolor}
\usepackage{fancyvrb}
\usepackage{booktabs}
\usepackage{inconsolata}
\usepackage{float}
\usepackage{breqn}
\usepackage[mathscr]{eucal}
\usepackage{subfigure}
\usepackage{fancyvrb}
\usepackage{todonotes}

\graphicspath{{img/}}

\newcommand{\rev}[1]{\textcolor{black}{#1}}
\newcommand{\ch}[1]{\textcolor{black}{#1}\xspace}

\newcommand{\new}[1]{\textcolor{black}{#1}\xspace}

\newcommand{\cmd}[1]{\xspace{\tt#1}\xspace} 
\newcommand{\name}[1]{\xspace{\sf#1}\xspace}  

\newcommand{\fogbrain}{\name{FogBrainX}}
\newcommand{\fogarm}{\name{FogArm}}
\newcommand{\core}{\fogarm \name{Core}}
\newcommand{\fogwatcher}{\name{FogWatcher}}
\newcommand{\fogmon}{FogMon\xspace}
\newcommand{\prolog}{Prolog\xspace}
\newcommand{\creasoning}{\textit{continuous reasoning}\xspace}

\newcommand{\requirements}{\texttt{requirements.yml}\xspace}
\newcommand{\compose}{\texttt{docker-compose.yml}\xspace}

\newcommand{\oq}{``}

\begin{document}
\title{Continuous QoS-compliant Orchestration\\ in the Cloud-Edge Continuum\thanks{Copyright \copyright\ 2022 for this paper by its authors. Use permitted under Creative Commons License Attribution 4.0 International (CC BY 4.0). Work partly supported by projects: \textit{Energy-aware management of software applications in Cloud-IoT ecosystems} (RIC2021PON\_A18), funded with ESF REACT-EU resources by the \textit{Italian Ministry of University and Research} through the \textit{PON Ricerca e Innovazione 2014--20}; \textit{O. Carlini Scholarships 2020} funded by the GARR Consortium; \textit{Including people in smart city applications} (PID2021-125527NB-I00), funded by the \textit{Spanish Ministry of Science and Innovation}.}
}
\titlerunning{Continuous QoS-compliant Orchestration in the Cloud-Edge Continuum}
%
\author{Giuseppe Bisicchia\inst{1}\inst{*} \and Stefano Forti\inst{1} \and Ernesto Pimentel\inst{2} \and Antonio Brogi\inst{1}}
\authorrunning{G. Bisicchia et al.}
%
\institute{Department of Computer Science, University of Pisa, Pisa, Italy
\and 
ITIS Software, University of Málaga, Málaga, Spain
\\
* Corresponding author: \email{giuseppe.bisicchia@phd.unipi.it}
}
\maketitle              
\begin{abstract}

%
The problem of managing multi-service applications on top of Cloud-Edge networks in a QoS-aware manner has been thoroughly studied in recent years from a decision-making perspective. However, only a few studies addressed the problem of actively enforcing such decisions while orchestrating multi-service applications and considering infrastructure and application variations.
In this article, we propose a next-gen orchestrator prototype based on Docker to achieve the continuous and QoS-compliant management of multiservice applications on top of geographically distributed Cloud-Edge resources, in continuity with CI/CD pipelines and infrastructure monitoring tools. Finally, we assess our proposal over a geographically distributed testbed across Italy.

\end{abstract}

\section{Introduction}
\label{sec:intro}

To support the growth of the Internet of Things (IoT) devices, new infrastructural architectures have been proposed, relying on computing, storage and, networking resources along the so-called Cloud-Edge continuum \cite{alloneneedstoknowfog}. Most of them -- e.g., Fog, Edge, Mist computing \cite{edgeintroduction,uehara2018mist} -- are based on the idea of employing computational capabilities closer to application end-users or, more generally, to data sources.
Such a continuum is characterised by its high dynamicity, device/connection heterogeneity, and availability of resources \cite{Svorobej2020}.

In contrast with the Cloud paradigm, the Cloud-Edge continuum can better support the deployment of next-gen IoT applications, usually featuring strict run-time
constraints on, for instance, required IoT devices, latencies, and bandwidth availability (e.g., virtual reality, remote surgery, online gaming)~\cite{bellavistafog}. Indeed, such applications are developed in the form of (possibly) hundreds interacting microservices, each of them with its own peculiar requirements.
As Cloud-Edge resources, also modern applications rapidly evolve over time, being continuously and collaboratively developed and released through automated tools in the \textit{Continuous Integration/Continuous Deployment} (CI/CD) pipelines~\cite{bobrovskis2018survey}.

In this context, much literature focused on determining the best QoS- and context-aware
placements\footnote{A \textit{placement} maps each managed microservice to a node of the infrastructure, in such a way all the application's QoS requirements are satisfied.} of multiservice IoT applications to Cloud-Edge infrastructures, by mainly exploiting search-based and mathematical programming solutions, e.g.,~\cite{brogi2019place}, \cite{herrera2023continuous}, \cite{Salaht2020AnOO}, and~\cite{MahmudApplicationManagement}. 
However, even if the compelling need for QoS-aware methodologies to place and manage application services onto Cloud-Edge infrastructure efficiently is evident \cite{velasquez2018fog,fogorch,costafogorch,7867735}, most existing proposals only referred to simulated environments due to the lack
of orchestration platforms capable of monitoring the needed QoS attributes, and to limited availability of Cloud-Edge testbeds \cite{smolka2022evaluation}. Particularly, the problem of designing platforms and methodologies for the orchestration and management of multiservice applications in a Cloud-Edge setting is a challenging one, having to deal with the scale and dynamicity of Cloud-Edge networks and of next-gen applications, but has only been marginally addressed.

In light of these needs, new solutions to support the QoS-aware orchestration and management of next-gen multiservice distributed applications suited for Cloud-Edge infrastructures could bring several benefits. To the best of our knowledge, none of the most popular orchestrators for managing digital infrastructures and services (e.g.,~Docker Swarm, Kubernetes) supports a continuous (i.e.,~incremental and differential) decision-making process that implements a scalable, QoS- and context-aware orchestration of microservices, ensuring suitable service placement and deployment on top of highly dynamic infrastructures, in continuity with the CI/CD pipeline and always (re-)considering the current infrastructure conditions.

In this article, we design and develop a next-gen prototype orchestrator\footnote{Freely available at: https://github.com/di-unipi-socc/FogArm}, \fogarm, to achieve the continuous and QoS-compliant management of multiservice applications on geographically distributed Cloud-Edge networks, in continuity with CI/CD pipelines and infrastructure monitoring tools. We also assess the performance of our orchestrator over a real-world, geographically distributed testbed.

\begin{figure}
    \centering
    \includegraphics[width=\textwidth,trim={0 6cm 0 6cm},clip]{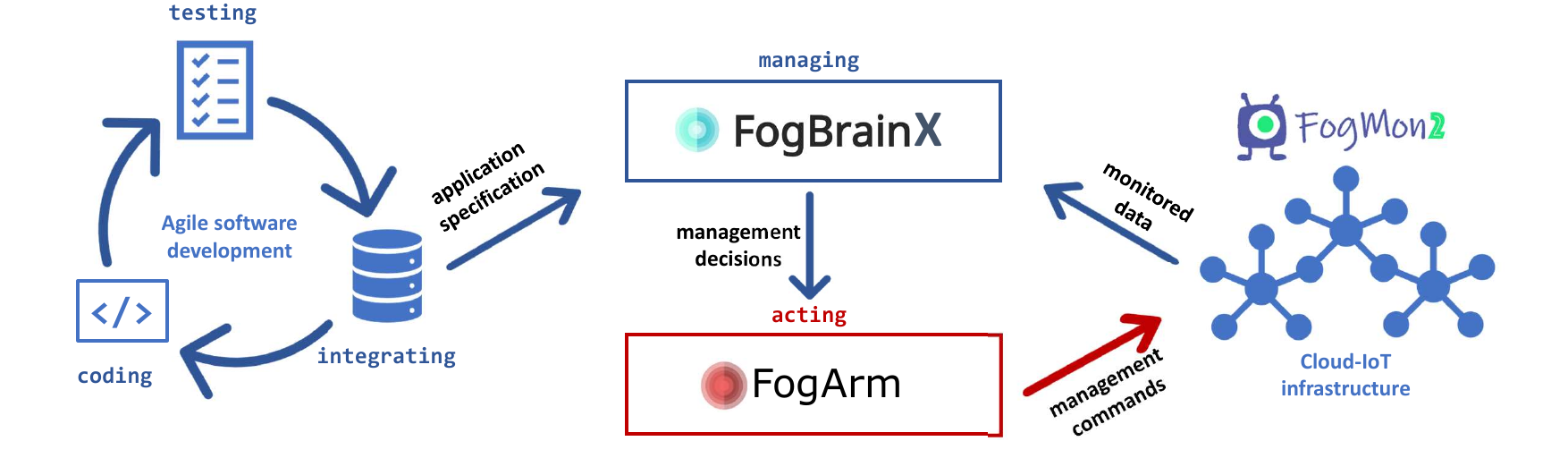}
    \caption{Bird’s-eye view of \fogarm.}
    \label{fig:birdsfax}
\end{figure}

As shown in Fig. \ref{fig:birdsfax}, \fogarm, strictly interacts with a monitoring tool (\fogmon in our case \cite{FORTI2020}), and with \fogbrain \cite{my_fogbrain}, a declarative \creasoning\footnote{By mainly considering the migration of services suffering due to such changes in the infrastructure \rev{or in the application}, \creasoning, permits, on one hand, scaling to larger sizes of the placement problem by incrementally solving smaller instances of such a problem, \rev{thus acting as a booster for existing placement strategies and reducing} the time needed to make informed \rev{management} decisions. On the other hand, it can reduce the number of management operations needed to adapt the current deployment to the new infrastructure conditions, by avoiding unnecessary service migrations.} engine to make informed service placement and migration decisions for next-gen multiservice applications. \fogbrain, through continuous (i.e., incremental, differential) reasoning, provably reduces the time needed to make management decisions when only part of running application deployment is affected by changes in the Cloud-Edge infrastructures (e.g., crash of a node hosting a service, degraded network QoS) or when the application itself changes (e.g., changed application requirements, addition or removal of application services).

To the best of our knowledge, \fogarm represents a first complete prototype of a next-gen orchestrator for the continuous QoS-compliant management of multiservice applications on top of geographically distributed Cloud-Edge infrastructures.

The rest of this article is organised as follows. Sect. \ref{sec:background} briefly presents the main peculiarities of the tools exploited by \fogarm, viz., \fogmon (Sect. \ref{sec:fogmon}) and \fogbrain (Sect. \ref{sec:fogbrain}). Then, Sect. \ref{sec:body} discusses the architecture and behaviour of \fogarm. Sect. \ref{sec:exps} illustrates the assessment of \fogarm over a real Cloud-Edge testbed. Finally, Sect. \ref{sec:related} briefly discuss related work and Sect. \ref{sec:conclusions} concludes the article by highlighting some possible directions for future work.
\section{Background}
\label{sec:background}

In this section, we briefly present \fogmon \cite{fogmon2019} (Sect. \ref{sec:fogmon}) and \fogbrain \cite{my_fogbrain} (Sect. \ref{sec:fogbrain}) that are integrated into \fogarm to equip it with monitoring and reasoning capabilities, respectively.

\subsection{FogMon: Lightweight Infrastructure Monitoring}
\label{sec:fogmon}

\fogmon is a TRL5 C++, distributed monitoring tool targeting Cloud-Edge computing settings\footnote{Available at https://github.com/di-unipi-socc/FogMon.}. \fogmon measures and statistically aggregates node capabilities (viz., CPU, RAM and HDD) as well as connected and available IoT devices and link QoS (viz., latency and bandwidth).
It leverages on a self-organising and self-restructuring peer-to-peer topology, that can run on any TCP/IP network, based on a two-tier \textit{Leader-Follower} architecture and gossiping protocols \cite{gossiping} for communicating among peers. 

\textit{Follower} agents have the task of monitoring the capabilities on their associated node. They are divided into groups, each group assigned to a specific \textit{Leader}.

Leaders perform all the tasks of a Follower and periodically aggregate data gathered by the Followers in their group. Furthermore, through gossiping, Leaders share among them the aggregated data collected from their Followers. Moreover, periodically, Leaders publish on a common endpoint a report containing all the gathered information both from their Followers and the other Leaders. The most recent report received is published as the current global report on the monitored infrastructure. The published reports and the communication between peers exploit JSON messages.

Leaders, also compute estimates of bandwidth and latency between Followers belonging to distinct groups. Indeed, Followers inside the same groups directly measure the link performance among them, but to avoid network congestions due to the exponential explosion of possible links between peers, QoS parameters between Followers in different groups are only estimated and not directly measured. In detail, Followers directly measure the network parameters only among Followers in the same group and with their Leader. Furthermore, Leaders directly measure the link performance among them. Thus, the QoS of a link between two Followers in a different group is estimated by composing the measurements between each Follower and its Leader and between the two Leaders.

Peers self-organise into an overlay peer-to-peer network constructed upon a proximity criterion based on latency distances among nodes. Indeed, any new node joins as a Follower and eventually selects its own Leader the one with the minimum measured latency. Periodically, during its activity, or after a failure, a node performs this procedure again to find the best suitable Leader. This approach is designed to face the high dynamicity of the Cloud-Edge continuum and to continuously adapt to a changing environment. For the same reason, also the role of Leader and Follower are dynamically assigned and can vary over time, restructuring the network topology by exploiting the k-medoids algorithm \cite{kmedoids}. Finally, the monitored data are also replicated by each Leader that, together with the eventual consistency of such data achieved through gossiping, make \fogmon capable of resisting the failure of some Leaders.

\fogmon shows a very low footprint in terms both of hardware and bandwidth resources, performing its probing tasks with low overhead. Furthermore, the two-tier peer-to-peer architecture avoids (e.g., due to node or link failures) a single point of failure as well as increases the scalability.
\fogmon is also released as a Docker image, thus being cross-platform on any Docker-compliant node.

\subsection{Declarative Continuous Reasoning in the Cloud-Edge Continuum}
\label{sec:fogbrain}

\fogbrain \cite{my_fogbrain} is a declarative \creasoning engine to make informed service placement and migration decisions for next-gen multiservice applications in Cloud-Edge settings. The idea behind the use of \creasoning is to mainly consider the migration of services in need of attention while preserving as much as possible the placement of the other services.

The \creasoning strategy employed by \fogbrain helps in reducing the time needed to make management decisions while only a part of application deployment is affected by infrastructural changes (e.g., traffic congestion, node failures) or triggers by a CI/CD pipeline (e.g., new service, requirements updates). 
Through \creasoning, we can scale our approach to larger applications and infrastructures by incrementally solving smaller instances of the placement problem and at the same time, reducing the number of management operations (e.g., avoiding unnecessary migrations).

\fogbrain reacts to changes in the infrastructure, in the application requirements and service addition and removal. \fogbrain is designed as a booster for existing placement strategies, being able to adapt to different approaches, improving their performance and reducing the size of the considered problem.

When \fogbrain is triggered, it verifies if a placement for an application already exists, if not, the default placement strategy (e.g., exhaustive search, heuristics) is applied to find a valid placement. Otherwise, the \creasoning methodology is performed. In this last case, \fogbrain first determines all services that have been added to the application from the latest commit and the services that have to be migrated (due to infrastructural or requirements changes). Such services are given in input to the default placement strategy to complete the partial placement of the services that have not been migrated, finding a new valid placement.

\fogbrain has been successfully applied through simulations over a lifelike small use case based on real data, and assessed at increasing infrastructure sizes and different variations rates up to thousands of nodes. It showed a speed-up of $50$ to $1000\times$ in terms of average inferences across different large-scale infrastructure sizes (i.e., from 320 to 1280 nodes).

\section{Design \& Implementation of \fogarm}
\label{sec:body}

In this section, we illustrate the architecture and functionalities of \fogarm, a next-generation orchestrator prototype designed to perform continuous and QoS-compliant management of multi-service applications on top of highly dynamic and geographically distributed resources such as the Cloud-Edge continuum. Sect. \ref{sec:arch} discusses the general component-wise architecture of \fogarm. Sect. \ref{sec:beh} illustrates the run-time behaviour of our orchestrator, highlighting the interactions of the components through three main scenarios. Finally, Sect. \ref{sect:impl} describes the actual implementation of \fogarm and its components.

\subsection{Architecture of \fogarm}
\label{sec:arch}

\fogarm enables:
\begin{itemize}
    \item the integration with CI/CD pipelines and infrastructure monitoring tools (e.g., \fogmon \cite{fogmon2019}),
    \item the execution of management decisions made by \fogbrain, and
    \item user interactions via a Web GUI and a Command Line Interface (CLI). 
\end{itemize}

\begin{figure}[!ht]
    \centering 
    \subfigure[Application page.]{\includegraphics[width=\textwidth]{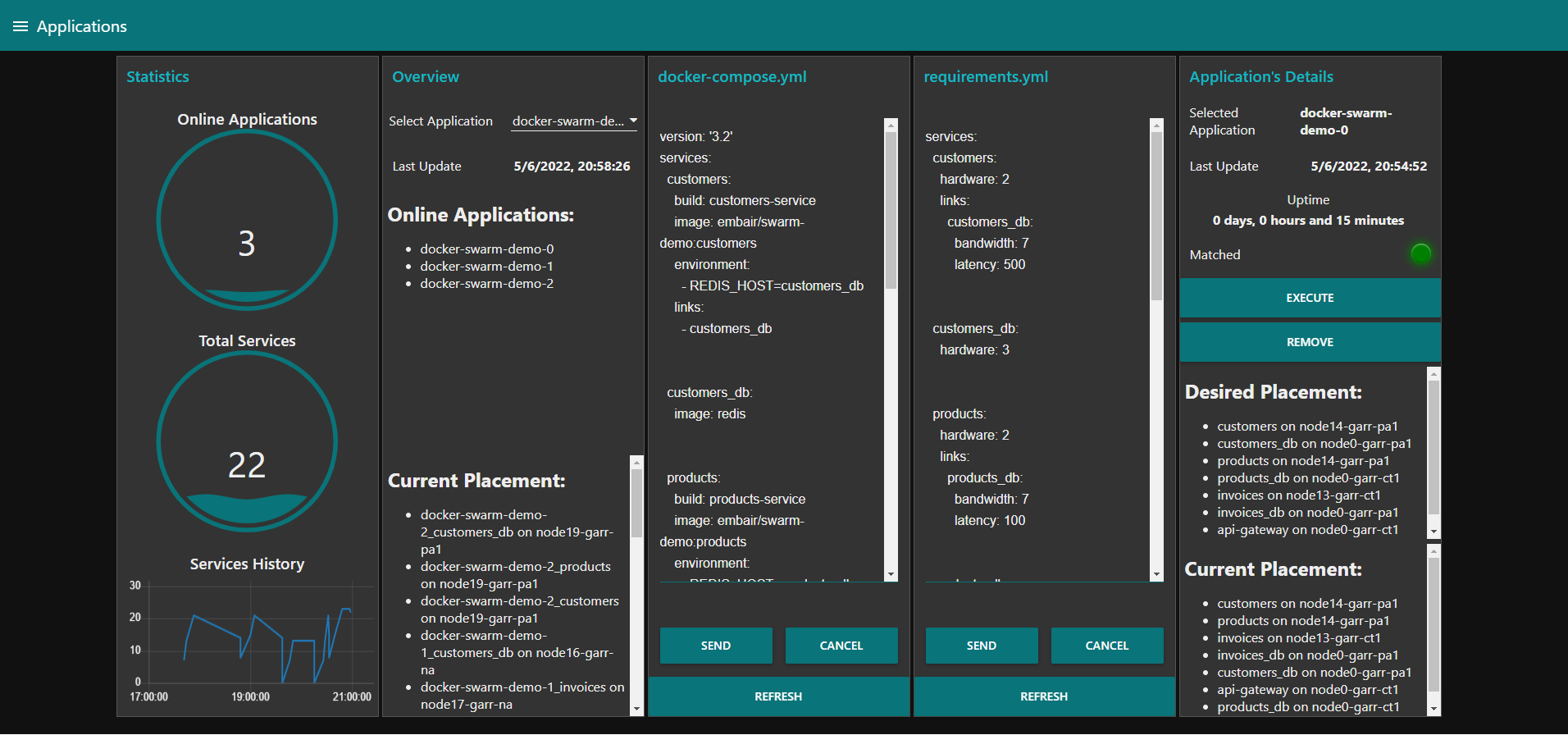}}
    \hfill
    \centering
    \subfigure[Node page.]{\includegraphics[width=\textwidth]{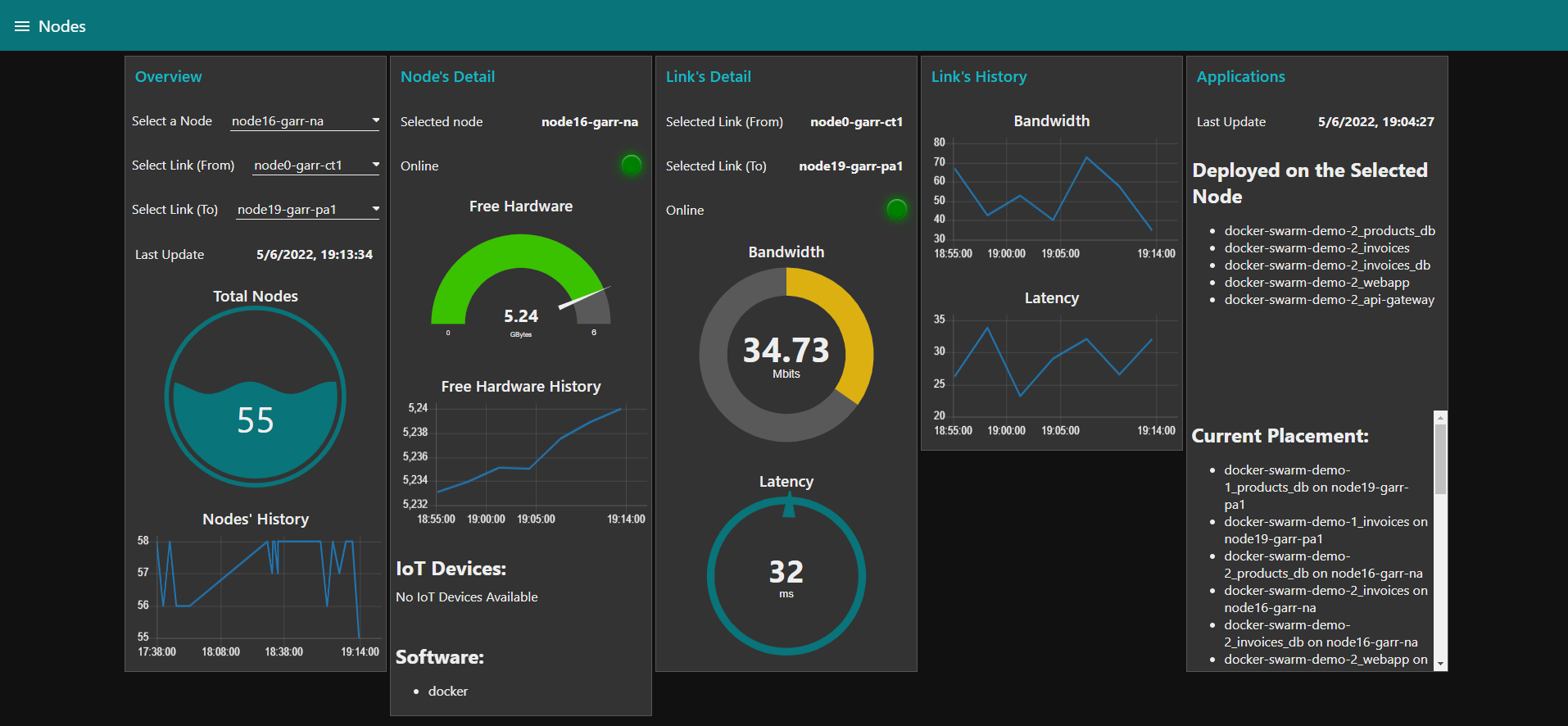}}
    \hfill
    \centering
    \caption{The WebGUI.}
    \label{fig:webgui}
\end{figure}

Fig. \ref{fig:blackboxfax} sketches the overall architecture of \fogarm, with the services that enable the above features. Namely:

\begin{description}
\item[FogArm Core,] which retrieves, in continuity with one or more CI/CD pipelines,  the information about the applications to be managed and the current state of the infrastructure, exploited by \fogbrain to determine management decisions. \fogarm then transforms such decisions into executable actions and implements them through Docker Swarm. It also offers a CLI through which users can interact by requesting the execution of actions and/or by monitoring the current state of the managed resources and applications. \core needs two files for each managed application, listed in Fig. \ref{fig:faxfiles}. The standard \texttt{docker-compose.yml} file (Fig. \ref{fig:faxfiles}a) contains information to configure the application’s services and the \texttt{requirements.yml} file (Fig. \ref{fig:faxfiles}b) describes the software, hardware and IoT devices requirements for each service reported in the \texttt{docker-compose.yml} file, as well as latency and bandwidth to other services. These requirements are used to automatically generate a suitable \prolog file exploited by \fogbrain as application specification.

\begin{figure}
    \centering
    \includegraphics[width=\textwidth,trim={0 4cm 0 4cm},clip]{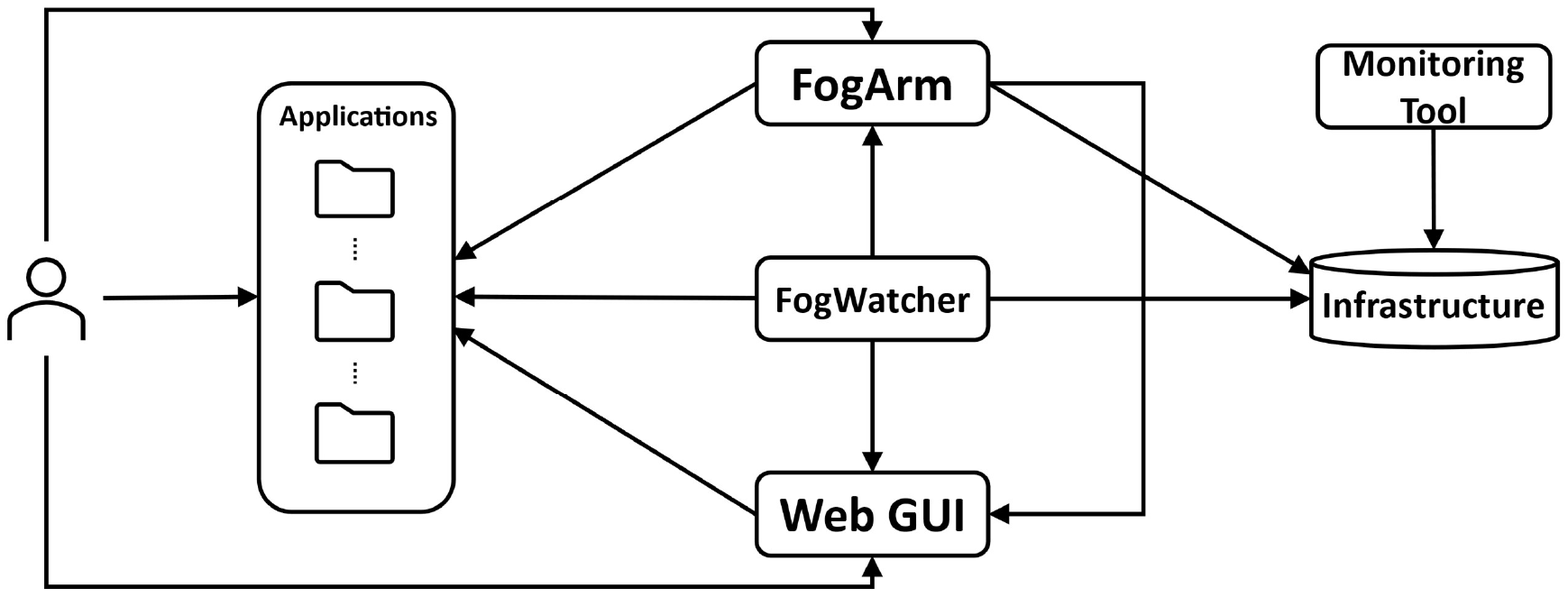}
    \caption{Architecture of \fogarm.}
    \label{fig:blackboxfax}
\end{figure}

\begin{figure}

\begin{minipage}[t]{.555\textwidth}
\begin{framed}
\begin{Verbatim}


version: "3.3"

services:
  web:
    image: localhost:5000/stackdemo
    build: .
    ports:
      - "8000:8000"
  redis:
    image: redis:alpine
    
\end{Verbatim}
\end{framed}
\centering(a) \texttt{docker-compose.yml}
\end{minipage}%
\hfill
\begin{minipage}[t]{.35\textwidth}
\begin{framed}
\begin{Verbatim}
services:
  redis:
    hardware: 6
    links:
      web:
        bandwidth: 20
        latency: 150
  web:
    hardware: 3
    links:
      redis:
        bandwidth: 50
        latency: 500
\end{Verbatim}
\end{framed}
\centering(b) \texttt{requirements.yml}
\end{minipage}
\caption{The application specification files.}
\label{fig:faxfiles}
\end{figure}

\item[FogWatcher,] which is a daemon service that monitors whether updates have occurred in the specification of managed applications or the status of the infrastructure. It checks whether the desired and current application placements do not match so as to trigger \core to enforce appropriate actions. Last, it checks user triggers coming from the Web GUI (e.g., updates on services' requirements). By, automatically triggering \core, \fogwatcher automatically guarantees the requirements of each application at run-time, without the need for human intervention. Thus, \fogwatcher allows closing the management loop of an application in an automated cycle starting from the CI/CD pipeline, passing through its deployment and any necessary migrations in the presence of infrastructural or application requirements changes.

\item[Monitoring Tool,] which takes care of retrieving the current state of all nodes and links in the considered infrastructure. This information is converted into a series of \prolog facts ready to be used by \fogbrain. 
During our experiments, we employed the \fogmon monitoring tool (Sect. \ref{sec:fogmon}). An adapter downloads the latest available \fogmon report and translates it into a set of Prolog facts monitored by \fogwatcher.

\item[Web GUI,] which shows all updates on the status of the infrastructure, applications' requirements, and their current and desired placement (Fig. \ref{fig:webgui}). It allows users to monitor the global or individual status of nodes and links, and to read and modify application specifications. It also offers the possibility to observe on which node the various services are currently placed and to manually request (and possibly actuate) the new placement for a given application, or to undeploy an application.  

\end{description}

\fogarm operates autonomically by constantly monitoring the state of the infrastructure, the managed applications and their placements. It fully exploits the incremental approach of \creasoning by reducing management operations only to those services in need of attention, as identified by \fogbrain. Furthermore, \fogarm is capable of simultaneously orchestrating several different multi-service applications on highly dynamic and geographically distributed infrastructures.

Note that the choice of the CI/CD pipelines and the monitoring tools are completely orthogonal and transparent to \fogarm. Indeed, \fogwatcher periodically checks if the information about the requirements of the applications and the state of the infrastructure has changed compared to the previous iteration and triggers \core when needed, independently from the entity (tool or human) that updates those state information. 

\fogarm leverages Docker Swarm at a low level to be able to focus on the innovative aspects of the orchestration process, delegating the basic mechanisms (e.g., deployment and allocation of resources) to a widely used and validated tool. We chose Docker Swarm for two main reasons. On one hand, it relies on Docker containers to feature flexibility and ease of use, since Docker containers are the \textit{de facto} standard to deploy microservices. On the other hand, it offers all the low-level features needed for the management of clusters and the deployment of services.

Overall, \fogarm translates \fogbrain's decisions into actions on containers by exploiting Docker's constraints \footnote{https://docs.docker.com/engine/reference/commandline/service\_update/}. Constraints, enable specifying that a given service must necessarily be deployed to a specific node (by its \texttt{hostname}). Note that when a constraint is specified (or modified), if the service is on the wrong node, Docker automatically takes care of migrating from the node where the service is located to the one requested in the constraint.

Thus, exploiting \fogbrain and Docker, \fogarm performs a complete monitor-analyse-plan-execute flow through the continuous monitoring of services' requirements and infrastructure status and, the interaction with Docker in Swarm mode, as depicted in Fig.~\ref{fig:birdsfax}.

\subsection{\fogarm's Behaviour}
\label{sec:beh}

\begin{figure}
    \centering
    \includegraphics[width=\textwidth,trim={0 3cm 0 3cm},clip]{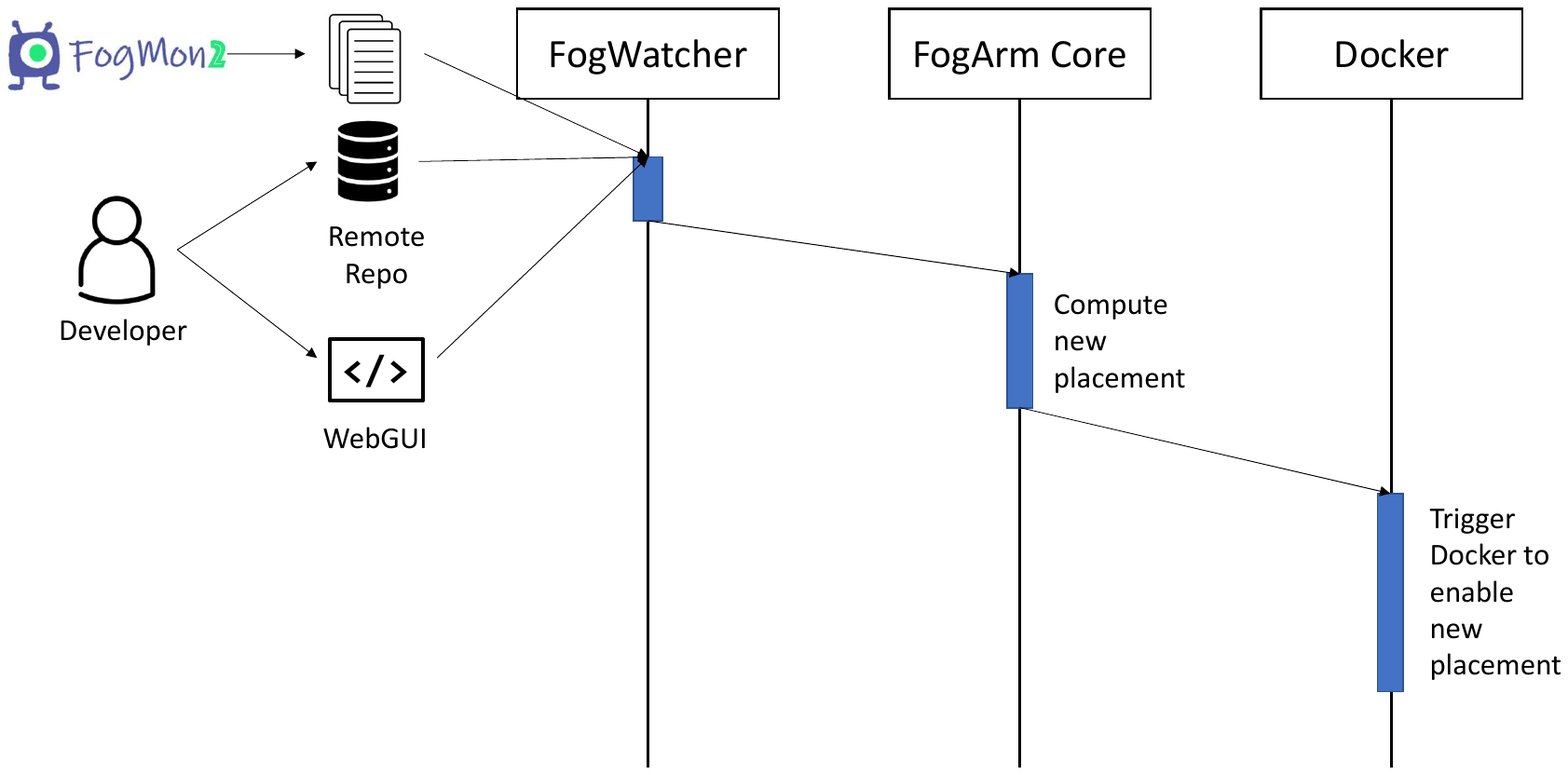}
    \caption{\fogarm's UML interaction diagram.}
    \label{fig:fogarmuml}
\end{figure}

In this section, we discuss and highlight the behaviour of \fogarm, in terms of the interactions of its components, illustrating three main scenarios viz., changes from the CI/CD pipeline, infrastructure changes and triggers from the Web GUI. These scenarios are sketched in Fig. \ref{fig:fogarmuml}.

\subsubsection{Changes from the CI/CD pipeline}

One of the most common scenarios working with \fogarm is when a new commit is received through the CI/CD pipeline. In this case, the remote application repository is updated with the new code and application requirements might change. Periodically, \fogwatcher checks whether changes have occurred in each managed repository. When a change is spotted, \fogwatcher triggers \core. \core collects the application requirements from the local repository (i.e., the \compose and the \requirements files) and the latest infrastructure report. This information is translated into a set of \prolog facts that are given as input to \fogbrain. Eventually, \fogbrain (possibly) outputs a new placement. Then \core, computes the differences between the current placement and the new one generated by \fogbrain. If the two differ, \fogarm generates a set of Docker commands to reconcile the actual placement with the desired one.

\subsubsection{Infrastructure Changes} Similarly, if a change occurred on the infrastructure resources, \fogmon publish it in the new report. Periodically, \fogwatcher verifies whether some changes have occurred in the infrastructure and triggers \core. The same procedure of the previous case is then performed, ending with the (possible) new sets of docker commands to reconcile the current placement into the new one determined by \fogbrain.

\subsubsection{Triggers from the Web GUI} Finally, the last main source of updates is the Web GUI. Indeed users, besides monitoring the status of the infrastructure and the managed application and services, can also change the \compose and \requirements files through the GUI. \fogwatcher periodically checks if new updates are published by the Web GUI. If so, first the updates are saved in the local repository and then \fogwatcher triggers \core as in the CI/CD pipeline scenario.

\subsection{\fogarm's Implementation}
\label{sect:impl}

In the following paragraphs, we briefly discuss the design and implementation of \fogarm components. Sect. \ref{sec:fax_backend} details the technological aspects of the backend components, while Sect. \ref{sec:fax_frontend} illustrates the frontend.

\subsubsection{BackEnd}
\label{sec:fax_backend}

\paragraph{FogArm Core}
\label{sec:fax_proto}

is the central component of the whole \fogarm architecture. It is implemented in Python3 exploiting the \textit{argparse} library to implement the CLI, the \textit{Docker SDK for Python} to interact with Docker and the \textit{PySwip} library, which together with a \prolog script, makes possible the interaction with \fogbrain. The main task of \core is to collect all the necessary information on the applications and the state of the infrastructure, consult \fogbrain and actually apply its management decisions.

The main way to interact with \core, and more generally with \fogarm, is through its CLI. Through it, it is possible to add, remove and manage applications, as well as interact with \fogwatcher and consult the status of managed applications and their placement.

More in detail, the main CLI commands of \fogarm are:

\begin{description}
\item[\texttt{add}] which deploys, for the first time, the application specified in the path argument, if entered, otherwise in the current folder.
\item[\texttt{exec}] which performs a reasoning step, of one application if specified, otherwise for all the applications, by verifying if the current placement is still valid and possibly carrying out the necessary operations to add, remove or migrate the application's services.
\item[\texttt{rm}] which removes one or all applications from the infrastructure.
\item[\texttt{status}] For each application it displays the desired placement, the current one and checks if the two match or not.
\item[\texttt{watcher}] which enables starting, stopping, or restarting \fogwatcher and displaying if it is running or not.
\end{description}

As for \cmd{add} and \cmd{exec}, the two commands perform rather similar functions, with the only difference that in the case of \cmd{add}, being the first deployment, some additional information is stored by the system for future use and in the case of \cmd{exec} it is possible to specify the application whose reasoning process the user wants to execute also by its name (or it is also possible to request the execution of all applications).

Once the application to orchestrate has been identified, by path or name, and after any additional information has been saved, \core proceeds by verifying the existence of \texttt{docker-compose.yml} and the possible existence of \texttt{requirements.yml}\footnote{With \texttt{requirements.yml} it possible to \oq annotate" each service reported in the \texttt{docker-compose.yml} file with quality and quantity requirements that the service needs to be able to correctly perform its tasks. The conversion into a set of \prolog facts is then a simple 1-to-1 mapping.}. Once the data have been retrieved, this information is used to generate a series of \prolog facts representing the application specifications.

These facts together with the \prolog file containing the updated status of the infrastructure are passed to \fogbrain, which checks whether a deployment already exists. If it exists, \fogbrain checks whether this is still valid, if not or if the deployment does not exist, \fogbrain is asked to generate a new placement, possibly by applying \creasoning.

Three lists are then generated from the computed deployment, comparing the placement obtained with the previous one. The services to be deployed (together with the relative chosen node, i.e., those services that are in the new placement but not in the old one), those to be removed (i.e., those services that are in the old placement but not in the new one) and those to be migrated from one node to another (i.e., those services that are in both placements but are assigned to different nodes). If a previous deployment does not yet exist, all services are considered to be added.

Once these three lists have been defined, \core takes care of actually executing \fogbrain's decisions by interacting with the Docker CLI.

As aforementioned, \core translates \fogbrain's decisions into actions on containers by exploiting Docker's constraints. Through constraints, it is possible to specify that a given service must necessarily be deployed in a specific node (specified in our case by the node's \cmd{hostname}). If, for instance, the service is deployed for the first time, the constraint is simply added through a suitable instantiation of the command

\begin{Verbatim}[fontfamily=zi4, fontsize=\footnotesize, frame=single, framesep=1mm, framerule=0.1pt, rulecolor=\color{gray}]
docker service update --constraint-add node.hostname==NodeId AppId_ServiceId
\end{Verbatim}
\noindent
where \cmd{NodeId} is the unique hostname of that node and \cmd{AppId\_ServiceId} is composed by the identifiers of the application (i.e., \cmd{AppId}) and of the service (i.e., \cmd{ServiceId}).

Otherwise, if the service was already deployed, but \fogbrain migrates the service to another node, the previous constraint is first removed and then a new one is added through the sequence of commands

\begin{Verbatim}[fontfamily=zi4, fontsize=\footnotesize, frame=single, framesep=1mm, framerule=0.1pt, rulecolor=\color{gray}]
docker service update --constraint-rm node.hostname==OldNodeId AppId_SId 
docker service update --constraint-add node.hostname==NewNodeId AppId_SId
\end{Verbatim}
\noindent
where \cmd{OldNodeId} is the unique hostname of the node where the service is already placed and \cmd{NewNodeId} is the unique hostname of the new node where to deploy the service \cmd{SId} of the application \cmd{AppId}.

When a constraint is specified (or updated), if the service is on the wrong node, Docker automatically takes care of migrating such service from the node where it is located to the one specified by the constraint. Similarly, it is possible to request the complete removal of a service through

\begin{Verbatim}[fontfamily=zi4, fontsize=\footnotesize, frame=single, framesep=1mm, framerule=0.1pt, rulecolor=\color{gray}]
docker service rm AppId_ServiceId
\end{Verbatim}
\noindent
In this way, \core can manage the deployment and migrations of multi-service applications.
When the removal of an entire application is requested through \cmd{rm}, the Docker CLI is once again exploited to remove all the services of a given application, through
\begin{Verbatim}[fontfamily=zi4, fontsize=\footnotesize, frame=single, framesep=1mm, framerule=0.1pt, rulecolor=\color{gray}]
docker stack rm AppId
\end{Verbatim}
\noindent
If instead, the \cmd{status} is requested, the current placement of all managed applications is extracted and, application by application it is compared with the one determined by \fogbrain.

Finally, through the \cmd{watcher} command it is possible to interact with \fogwatcher being able to start, stop, or restart it on request.

\paragraph{FogWatcher}
\label{sec:fogwatcher}

is developed in Python3 exploiting the \textit{timeloop} library to implement periodic checks. The main task of \fogwatcher is to periodically monitor whether updates have occurred in the infrastructure or the specifications of the applications managed, triggering \core to perform the actions necessary to guarantee the desired QoS for each service, without the need for human intervention. 


\fogwatcher periodically monitors, with independent and customisable periods\footnote{Check periods, as well as other parameters of \fogarm's components, can be customised through a global \textit{configuration file}, updatable on the fly.}, four possible sources that may require \core activation:

\paragraph{The CI/CD pipeline} For each managed app, it is periodically checked whether the \compose or the \requirements~files have changed. For each of the files, a hash (exploiting the HASH256 function) of its content is computed. If the final hash does not correspond to the previous one, then a change has happened and \core is invoked to carry out a reasoning step through the \cmd{exec} command. In this way, \core, by invoking \fogbrain, checks for the application whether the required requirements are still satisfied after the specification change and, if not, carries out the necessary operations to add, remove and migrate the services.

\paragraph{The infrastructure} Furthermore, also the file containing the updated status of the infrastructure is monitored. \fogwatcher takes care of calculating the hash of that file and if the calculated hash does not correspond to the previous one, then for all the managed applications it is required to carry out a reasoning step.

\paragraph{The current placement} \fogwatcher also periodically checks, for each application, whether the desired placement and the current one correspond. If this is not the case, then the deployment computed by \fogbrain is removed and \core is invoked. Indeed, in a real system it may happen that after a node or a Docker error, or if there has been a manual intervention on one or more services, the current placement no longer corresponds to that requested by \fogbrain.

\paragraph{Live changes} Finally, \fogwatcher regularly queries the Web GUI to check if the user has requested any operation, e.g., removal of an application, execution of a reasoning step, update of the compose or requirements file, in which case the request is fulfilled and eventually a reasoning step is carried out.

\paragraph{Integration with \fogmon}
\label{sec:fax_fogmon}

In our implementation, \fogmon allows \fogarm to be always updated on the state of the infrastructure and to make informed management decisions to comply with application requirements.


After installing a \fogmon agent in each node of the infrastructure, such a tool periodically reports the updated status of the infrastructure in the form of a JSON file. Such monitoring data is made available through a REST API taken from \cite{new}.

Then \fogarm takes care to periodically consult that endpoint and obtain the latest available report. It checks whether the report obtained is different from the previous one\footnote{Note that, thanks to the \cmd{sensitivity} configurable parameter (i.e., threshold relative difference on average and variance to send differential reports) of the \fogmon agents, only variations that exceed the threshold level are reported. Thus, small fluctuations will not result in changes in the report. Hence, a change in the report implies the presence of at least one significant infrastructural change, which therefore needs attention.} (by comparing the hash of the two reports). If so, the JSON report is mapped into a set of \prolog facts and the input file is updated.

It will be the task of \fogwatcher to note that an infrastructural change has happened and to invoke \core to verify if service management operations are needed.

\subsubsection{FrontEnd}
\label{sec:fax_frontend}

\paragraph{The Web GUI}
\label{sec:fax_gui}

It is the main interface with which the user can interact with \fogarm. On one hand, the Web GUI offers a higher-level control of applications than the CLI. On the other hand, it enables to simply have an overall view of the entire system at a glance, with the possibility of paying close attention to the characteristics and properties of a single application or specific node and/or link.

The Web GUI is therefore designed to be primarily a monitoring interface with which to observe the actual status and the evolution of the system and be able to pay attention to detailed aspects, but which still offers the main functions for interacting with \fogarm and its managed applications.

The Web GUI is implemented through Node-RED\footnote{https://nodered.org/}, a flow-based development tool built on Node.js, and it is divided into two main pages, Application and Nodes. The former focuses on the global status of applications and services, with the ability to analyse a single application. The latter allows users to observe the overall state of the infrastructure and study a single node and/or a single link. 

We illustrate in more detail such two views of the Web GUI:

\paragraph{Applications}
\label{sect:applicationsgui}

The Applications page (Fig. \ref{fig:webgui}b), allows the users to monitor the overall state of the managed applications as well as analyse in detail the state of a particular application and interact with it, also offering the possibility to view and update its \compose and \requirements files. 

The page is divided into five panels.

\paragraph{Statistics} It allows having some global statistics on the current state of managed applications. In detail, it displays the total number of applications and services currently deployed as well as how the number of deployed services has varied over time.

\paragraph{Overview} This panel displays which applications are currently deployed and the current placement of each service belonging to those applications (i.e., on which node each service is deployed). It also reports when the last update was received and allows one to select a particular application to focus on.

\paragraph{docker-compose.yml \& requirements.yml} They show the last known state of the two related files of the application selected in the previous panel and offer the possibility to modify them live and send the changes to be implemented. Sending a change also automatically activates the execution of a reasoning step. If, on the other hand, a change has been applied but not yet sent, with a refresh button it is possible to return to the unmodified version. Cancel instead deletes all the contents allowing to rewrite the file from scratch.

\paragraph{Application's Detail} Shows various information about the application selected in the Overview panel. In particular, it displays when the last update for that particular application was received, the uptime and, for each service of that application, the desired and actual placement. A LED allows checking at a glance whether the two placements match (green light) or if they differ (red light). It is also possible to request the execution of a reasoning step for that application or remove it from the infrastructure.

\paragraph{Nodes}
\label{sect:nodesgui}

The Nodes page (Fig. \ref{fig:webgui}b) allows monitoring the current state of the whole infrastructure, offering also the users the possibility to focus their analysis on a specific node and/or link.

The Nodes page is divided into five panels.

\paragraph{Overview} Displays when it received the last update, the current number of available nodes and the evolution of that number over time. Allows also the users to select a particular node (among all the nodes of the infrastructure, not only the currently available ones) and/or link, specifying the two endpoints of the link.

\paragraph{Node's Detail} Shows if the selected node is online and its last knows status, in terms of available free hardware (i.e., RAM) and its evolution over time, and the available IoT devices and software (if any).

\paragraph{Link's Detail} Displays if the selected link is actually available and the last known value of the available bandwidth and latency.

\paragraph{Link's History} Allows the users to view how the bandwidth and latency of the selected link are changed over time.

\paragraph{Applications} Shows when the last updates for the applications are received and the current placement of the whole available services of all the deployed applications. If a node is selected, it also displays which services are deployed on that specific node (if any).
\section{Experimental Assessment}
\label{sec:exps}

In this section, we first illustrate the experimental assessment\footnote{We run all experiments over Virtual Machines (VMs) featuring 1 vCPU and 6GB of RAM, and running Ubuntu 20.04.3 LTS, provided by the GARR Consortium and spread across 3 regions (viz., Catania, Palermo and Turin). Nodes run Python 3.8.10, Docker 20.10.12, docker-compose 1.25.0 and SWI-Prolog 8.4.2 to support the correct execution of \fogbrain and \fogarm. A node in the Catania region is chosen as the leader from which the orchestration process is actually executed, interacting with the \fogarm CLI and the Web GUI.} of \fogarm at varying infrastructure sizes and number of applications over a real-world testbed (Sect. \ref{sec:scalabilityfax}). We then show how \creasoning can boost decision-making and application management times in a testbed made of 60 nodes and running 400 services (Sect. \ref{sec:crassessfax}).

In each experiment, the infrastructure is built and configured through an automatic script to guarantee the reproducibility of the procedure and results.

\subsection{Scalability assessment}
\label{sec:scalabilityfax}

We perform a scalability assessment of \fogarm to evaluate how scaling the size of the infrastructure and the number of managed applications and services might affect the orchestrators' performance.

\paragraph{Experimental Setup.}

\begin{figure}
    \centering
    \includegraphics[width=0.65\textwidth,trim={0cm 2.5cm 0cm 0cm},clip]{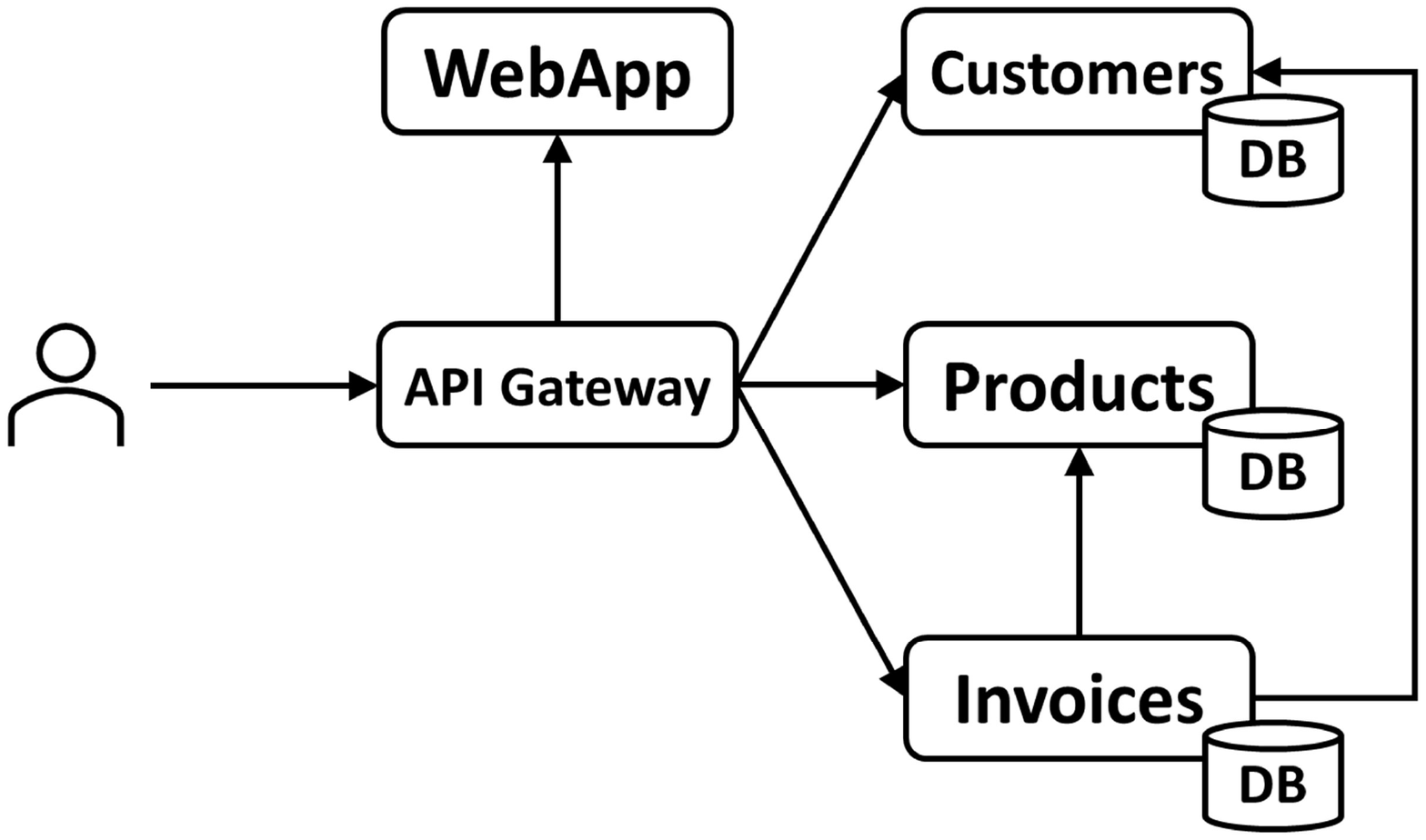}
    \caption{Example application.}
    \label{fig:appdemo}
\end{figure}

The assessment is divided into three experiments at increasing size of the infrastructure (viz. 15, 30 and 60 nodes) and number of application\footnote{Available at: https://github.com/michal-bures/docker-swarm-demo} replicas (viz. 10, 25 and 50). Each composed of 8 services (viz. a total of 80, 200 and 400 services, respectively).
Testbeds span 3 regions spread across Italy (viz., Catania, Palermo and Turin), with nodes evenly distributed among each region. Each node hosts a \fogmon agent to monitor available resources\footnote{In detail, whenever a new \fogmon report is received, we artificially reduce the available node's RAM of a value picked at random from a Gaussian distribution centred at 750MB with a standard deviation of 375MB. For each link, we artificially increase the latency by adding a  random value from a Gaussian distribution centred at 50 ms and with a standard deviation of 25 ms. Similarly, the bandwidth is artificially reduced by a value picked at random from a Gaussian distribution centred at 12.5\% of the available bandwidth and a standard deviation of 6.25\%. Last, nodes and links have a failure probability of 5\%.}.
We use replicas of the \oq Docker Swarm Demo" application\footnote{ https://github.com/michal-bures/docker-swarm-demo}, composed of 8 services (Fig. \ref{fig:appdemo}). For each replica, we simulate changes through the CI/CD pipeline, by randomly producing a new commit and/or updated service requirements\footnote{Each service can require from 250MB to 750MB of available RAM. For each service-to-service communication a latency from 200ms to 750ms and an available bandwidth from 10Mbits/s to 30Mbits/s. Furthermore, each service has a different probability, from 75\% to 100\%, of being added to the last generated commit, so to also experiment with the addition and/or removal of services at run-time.}.
Finally, we run and monitor experiments for 5 hours after the initial deployment of all applications.

\begin{figure}
    \centering 
    \subfigure[Average times.]{\includegraphics[width=0.48\textwidth]{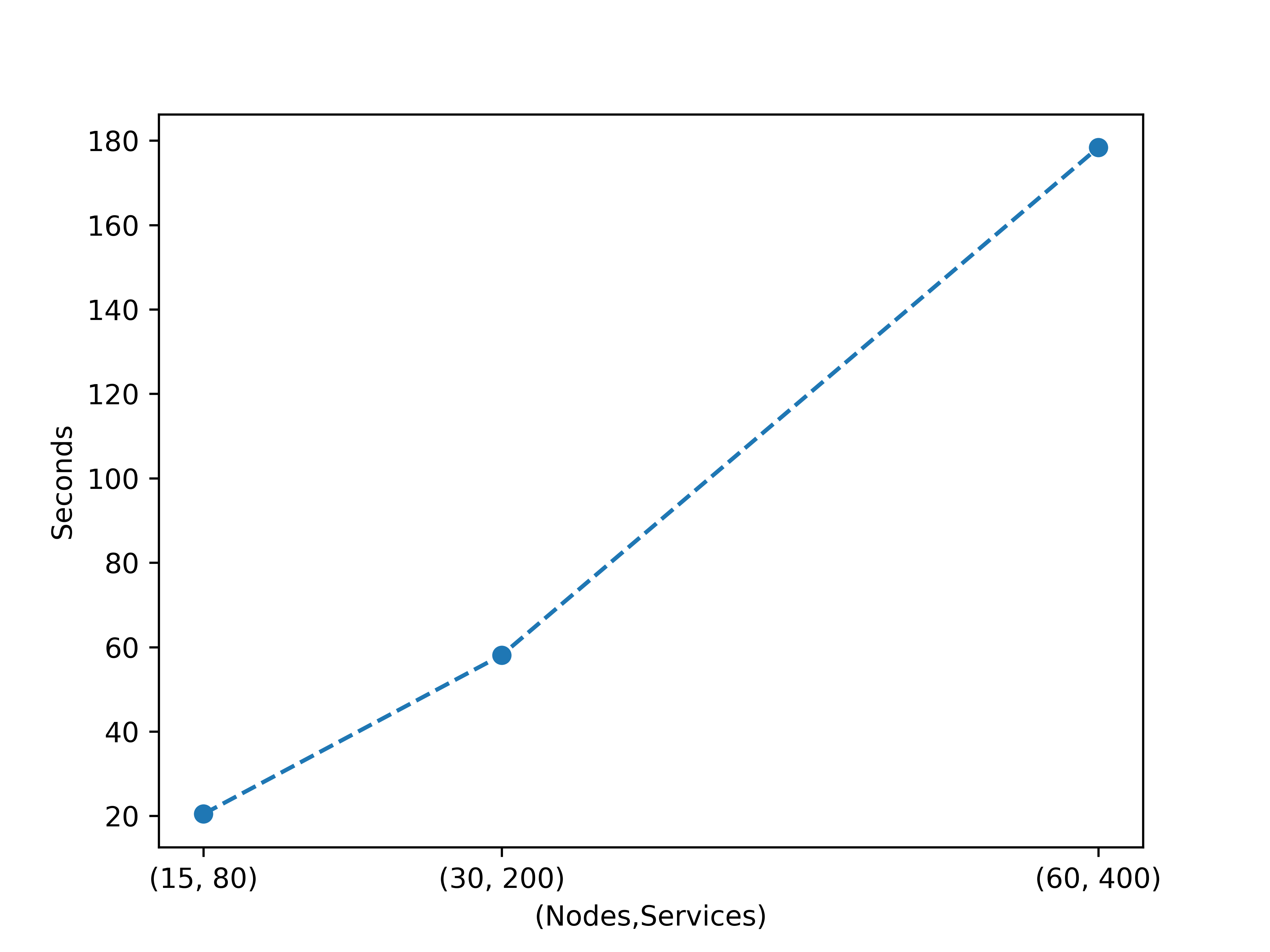}}
    \hfill
    \centering
    \subfigure[Average migrations.]{\includegraphics[width=0.48\textwidth]{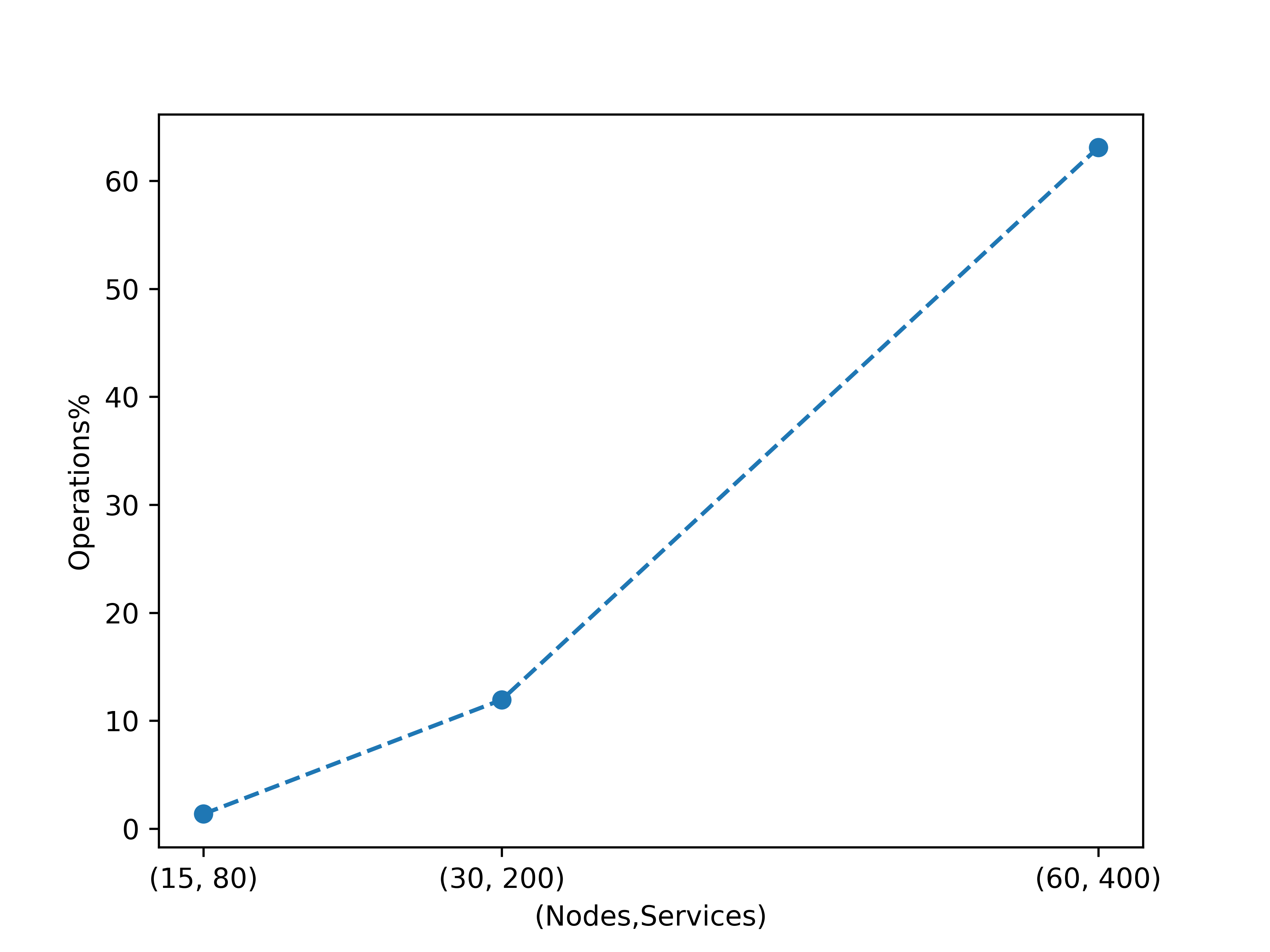}}
    \hfill
    \centering
    \caption{\ch{Results for the scalability assessment.}}
    \label{fig:scalexp}
\end{figure}

\paragraph{Experimental Results.} 

Fig. \ref{fig:scalexp}a illustrates the experimental results in terms of the average \fogarm execution times at increasing sizes of the infrastructure and number of managed services. Execution times sum up both the time needed by \fogbrain to take management decisions and by \fogarm to actuate such decisions by interacting with Docker Swarm. Fig. \ref{fig:scalexp}b shows the average percentage of migrations per execution step over the total number of services that could be migrated (i.e., thus excluding from the considered set of services those that are about to be added or removed).

Execution times (Fig. \ref{fig:scalexp}a) increase as the scale of the experiments increases. We experience an average execution time of 20 seconds when managing 80 services on 15 nodes (i.e., around 250 ms on average for each service). When managing 400 services, instead we experience an average execution time of 180 seconds (i.e., around 450 ms per each managed service). We have, then, an increase of less than 2$\times$ on the average execution times, while increasing by 5$\times$ the number of services and 4$\times$ the number of nodes. This behaviour relates to the variation of the average amount of required migrations over the experiments. Indeed, in smaller scenarios, we experience near-to-zero migrations, while we reach more than 60\% average migrations per execution step in larger scenarios (Fig. \ref{fig:scalexp}b).

\new{Overall, we observe an exponential increase both in execution times and migrations. Indeed, as the size of the infrastructure and the number of managed services increase, more resources are required by \fogmon to monitor the infrastructure state and by Docker to enact management decisions. Additionally, the larger number of managed services leads to a more dynamic system that, therefore, requires more migrations to maintain its optimal state. When the number of services and nodes increases, the chances of resource congestion and failures naturally increase and cause more migrations to satisfy application requirements. These results further highlight the need for efficient resource management strategies that can scale with the size of the infrastructure and the number of managed services.}

\subsection{Continuous Reasoning assessment}
\label{sec:crassessfax}

In this section, we compare the \creasoning approach of \fogarm against a version exploiting exhaustive search, i.e., possibly migrating services that are not affected by infrastructure changes or CI/CD triggers.

\paragraph{Experimental Setup.}

We consider the same settings of the largest scenario of the scalability assessment (i.e. 60 nodes evenly spread in 3 regions across the Italian national territory and 50 applications with most 8 services, for an overall number of 400 services). We also employ the same approach for generating application commits and changing infrastructure capabilities. Also in this case experiments last 5 hours.

\paragraph{Experimental Results.}

\begin{figure}
    \centering 
    \subfigure[Average times.]{\includegraphics[width=0.48\textwidth]{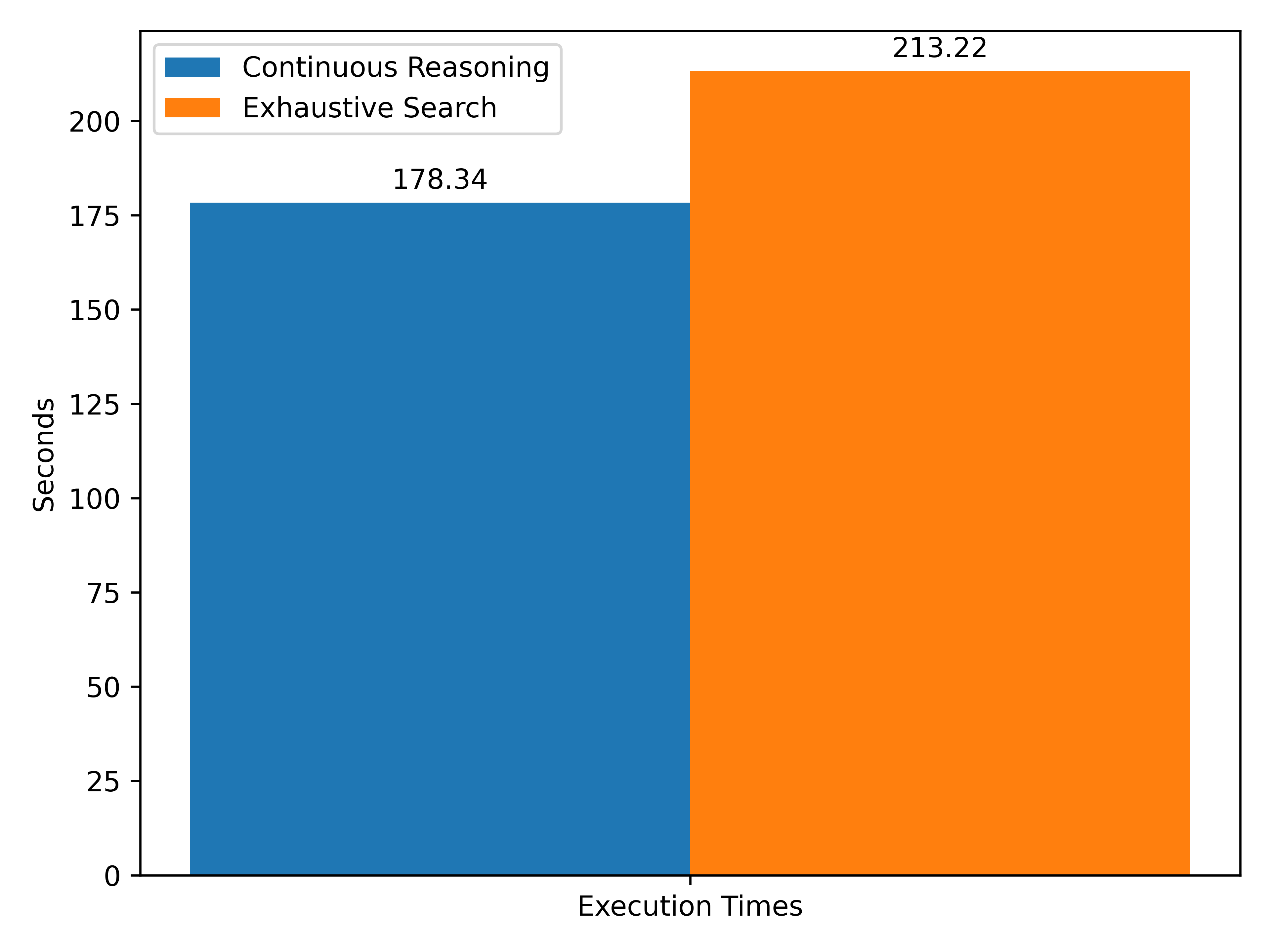}}
    \hfill
    \centering
    \subfigure[Average migrations.]{\includegraphics[width=0.48\textwidth]{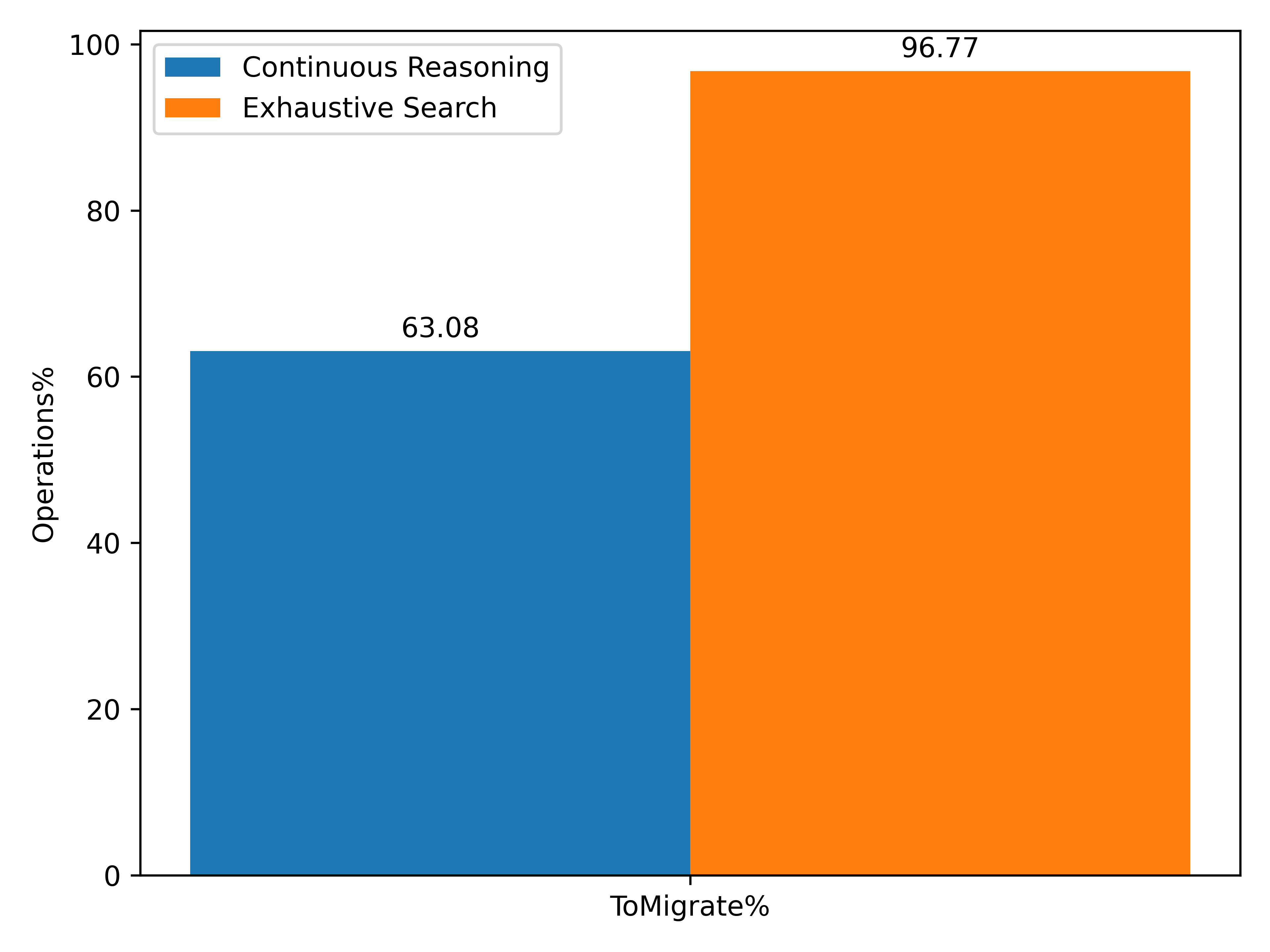}}
    \hfill
    \centering
    \caption{\ch{Results for the continuous reasoning assessment.}}
    \label{fig:crexp}
\end{figure}

Fig. \ref{fig:crexp}a compares the average execution times featured by \fogarm exploiting the \creasoning and the exhaustive search. The \creasoning strategy allows the orchestrator to save 35 seconds on average (i.e., around 15\%) while determining and enforcing a given placement in comparison with the exhaustive search. We can relate such improvement with the different number of migrations performed on average by the two strategies. Indeed, the exhaustive search enforces, on average, 33\% more migrations, increasing also the execution time required to enforce a given placement. Such results can be explained considering that \fogbrain, and hence \fogarm, tries to preserve the current deployment as much as possible. \fogarm, therefore, migrates only those services in need of attention due to infrastructure changes or CI/CD triggers. 

On the other hand, the exhaustive search does not pay any attention to preserving the current deployment, limiting itself only to finding a placement among all those admissible and possibly leading to migrating more services than necessary. \new{Overall, the results indicate that the \creasoning strategy is more efficient than the exhaustive search strategy in terms of execution time because it performs fewer migrations on average. Indeed, the \creasoning approach is designed to take into account and preserve the current deployment as much as possible while minimising the number of migrations required to achieve the desired placement.}

Finally, note that the exhaustive search is not QoS-aware, so we have no guarantees that the placement found meets the given requirements. \fogarm, instead, is designed to find only QoS-aware placements and to modify such placements when a CI/CD trigger or an infrastructure change occurs.
\section{Related Work}
\label{sec:related}

The problem of designing platforms and methodologies for the orchestration and management of multi-service applications in the Cloud-Edge continuum is a very well-known problem \cite{costafogorch,velasquez2018fog,7867735,fogorch}. The main difficulties are given by the scale, heterogeneity and diversity of the node's infrastructures. Another important factor is the high dynamicity, in terms of resource capabilities variation (e.g., memory and bandwidth), failures of nodes and links and, devices or users distribution, with their possible movements \cite{Svorobej2020}. Additionally, applications are composed of several different and heterogeneous software components with possible dependencies among them~\cite{barika2019orchestrating}.

Addressing these issues, \cite{fitor} proposes a device-aware, greedy approach to incrementally build service provisioning solutions. The authors divided infrastructure into two layers (viz., Fog and End devices), with the Fog layers subdivided into two sublayers (viz., High Fog and Mist). Considering such structure, the process iteratively decomposes an application into sub-components and greedily places each component considering its requirements. The process is repeated until all solution components are provisioned. \new{With respect to \fogarm, in \cite{fitor} a structural division of the infrastructure is proposed, while in our work we consider infrastructure as a graph of nodes. Indeed, we believe that a plain representation fits better the heterogeneity and pervasiveness of the Cloud-Edge continuum.}

Still dividing the Fog level, \cite{sorts} proposes a hybrid choreography/orchestration hierarchical strategy for service management in Fog environments, dividing the infrastructure into three layers (viz., IoT, Fog and Cloud). At the IoT level devices are organised in virtual clusters supporting the possibility of mobile devices. At such a level the management devices cooperate among them, in a choreography fashion, offering low response time and high resilience in presence of a device movement (i.e., shift into a different virtual cluster). At the Fog level, both strategies are performed depending on whether the fog device is in the south (i.e., closer to the IoT) or north (i.e., closer to the Cloud) region.  Choreography is performed in the south region, orchestration in the north region and on the Cloud. The three-level architecture offers higher dynamism to the lower levels while keeping a global view of the higher levels, in which possible optimisation is performed. \new{As discussed for \cite{fitor} also here a structural representation of the infrastructure is proposed. Furthermore, it is not clear how the services' requirements are managed and checked by their \textit{Resource Manager}. Finally, no prototypes are implemented for this work and there is also a lack of experimental results, instead, we proposed and assessed a fully automatic orchestrator, also gathering experimental results in an actual infrastructure.}

In \cite{chariot}, an orchestration middleware for IoT systems is presented. The orchestrator features a three-level architecture. Each layer, from the topmost to the bottom, comprises, namely, a system description language to describe fog infrastructures and services, persistent data storage and a management engine to formulate constraints that encode system properties and requirements in the form of satisfiability modulo theory (SMT). Such an engine enables the use of SMT solvers to determine a valid configuration at run-time. \new{Despite, the description language proposed is powerful and flexible it requires a greater effort by the developers with respect to that required by \fogarm in compiling the \requirements file and in \cite{chariot} users are required to compile also a description of the possible nodes' templates while in \fogarm the management of the nodes is fully transparent and automated. Finally, in the proposed version only the nodes' failures are accounted for, on the contrary, \fogarm is capable of managing nodes, networks and services failures.}

Working on Osmotic computing, \cite{osmoorch} discusses an orchestration architecture in which managed IoT applications, deployed in distributed environments, are modelled as a graph of MicroELements (MELs). A MELs graph models microservices, which implement specific functionalities, as well as microdata, representing information flows from/to IoT devices. The proposed orchestrators, through a Deep Learning process, generates MELs deployments based on previous experiences and eventually execute the obtained deployments manifest. \new{However, this work is only a theoretical work so neither a working prototype nor an experimental assessment is proposed. Furthermore, using a Deep Learning process usually decreases the explainability of the orchestrator, making it difficult to understand why a certain management decision was taken. Instead, the engine of \fogarm, \fogbrain, thanks to its declarative nature is explainable and it is possible to trace all the decision steps performed.} A model-driven approach is also exploited in \cite{mdd}, which proposes an attribute-driven framework for application development and service orchestration, assisting developers through the entire development lifecycle through a set of formal rules.

Exploiting Software Defined Networking, \cite{ctoso} presents a service orchestration mechanism to meet the latency and reliability requirements of IoT applications. Through a target optimisation function, a differentiated task offloading strategy is applied considering task attributes as well as communication and computation energy consumption and pre-estimated task offloading costs. Similarly, the problem of the placement of Virtual Network Function is studied in \cite{markovvfn}, through a multi-objective optimisation problem model that is converted to a problem which is solved by a Markov approximation technique.

Moving on to industry tools, the most popular solutions are based on container orchestration. In this field, the orchestrator manages the entire lifecycle of a container ranging from its creation to its destruction or termination and scaling or migrating containers if needed. Among the most widely used we have Docker in its Swarm mode\footnote{https://docs.docker.com/engine/swarm/} and Kubernetes\footnote{https://kubernetes.io/}, targeting clusters and datacenters. Both solutions do not offer high awareness of the services' requirements. Docker Swarm offers a system of constraints based on the labelling of nodes and services (i.e. placing a service if its labels match those of the nodes), while Kubernetes allows specifying simple CPU, memory and storage constraints. At the same time, these solutions are carried out manually, thus not coping with the high dynamism of Cloud-Edge infrastructures.

Finally, Topology and Orchestration Specification for Cloud Applications (TOSCA) \cite{tosca} is one of the first and main proposals for standardising the service orchestration in an extensible and flexible way \cite{bellendorf2018cloud,luzar2020examination}. TOSCA is an open-source language to define an interoperable model of cloud applications. TOSCA describes the components as well as the relationship and dependencies between them and their requirements and capabilities, thus enabling portability and automated management. With TOSCA applications are described as a typed, direct topology graph, representing components as nodes and the dependencies between them as links. For each component is also possible to describe its requirements as well as the needed operations and policies.

In \cite{tosker}, the TOSCA standard and the Docker ecosystem are exploited to propose an orchestration strategy for the management of multi-component applications based on a TOSCA-based representation. The approach allows specifying software components and Docker containers to form an application and automatically deploy and manage such applications. \new{With respect to \fogarm in this work the management of the applications is performed in a single machine and not on a distributed infrastructure. Furthermore, the orchestrator requires already developed management plans and it is not capable of finding the placement automatically.} On the contrary, \cite{app9010191}, proposes a TOSCA-based orchestration tool for automating the process of federating Kubernetes container clusters even across different cloud providers. \new{However, both the orchestrator illustrated do not support the connection with a CI/CD pipeline, and thus the modification of the applications' topology or requirements at runtime.}

Concluding, to the best of our knowledge, none of the existing orchestration solutions, unlike \fogarm, supports \creasoning or more generally a continuous (i.e., incremental and differential) scheduling process that makes QoS- and context-aware management of microservices, possibly ensuring the optimisation of the allocation of services on highly dynamic infrastructures, in continuity with the CI/CD pipeline. Furthermore, most existing proposals only referred to simulated environments due to the lack of orchestration platforms capable of monitoring the needed QoS attributes, and to the limited availability of Cloud-Edge testbeds~\cite{smolka2022evaluation}.

\fogarm, instead, is capable of autonomously adapting the deployment of the application in response to changes to the application specification coming from the CI/CD pipeline and to variations infrastructural detected through a distributed monitoring tool. When triggered \fogarm applies a \creasoning approach, through the interaction with \fogbrain. Finally, \fogarm is assessed in an actual geographically distributed infrastructure over the Italian national territory.

\section{Conclusions}
\label{sec:conclusions}

In this paper, we proposed \fogarm, a next-gen orchestrator prototype that performs fully automated and QoS-compliant continuous management of multiservice applications on top of highly dynamic and geographically distributed infrastructures.

To perform the orchestration process \fogarm interacts with different tools viz., \fogmon to gather the current status of the infrastructure's resources, \fogbrain to exploit its \creasoning approach to find a valid placement in a continuous and scalable way and finally, Docker Swarm to implement the low-level operations through Docker's constraints.

\fogarm continuously monitors the status of the infrastructure, the application's requirements and the current deployments searching for changes. When a change occurs, \fogarm verifies through \fogbrain if a new placement is required. If so, \fogarm generates the suitable management operations to accomplish the desired placements and interacts with Docker to perform the operations. 

Through \fogarm, developers are required only to define the requirements of each application's service. The whole process of deployment and management is fully automated without any required user action.
\fogarm works also in continuity with the CI/CD pipelines, supporting the current iterative and incremental development process. Additionally, users can interact with \fogarm through a CLI or a Web GUI.

Our experiments have shown how \fogarm can scale even on high dynamic, geographically distributed infrastructures with up to 60 nodes spread across Italy while managing up to 400 services from 50 applications, and continuously interacting with the CI/CD pipelines. Furthermore, the \creasoning methodology proved to save more than 15\% of the execution time (i.e., around 35 seconds) while migrating on average 33\% services fewer than the version of \fogarm featuring only the exhaustive search strategy.

To the best of our knowledge, \fogarm represents a first complete prototype of a next-gen orchestrator for the continuous QoS-compliant management of multi-service applications on geographically distributed Cloud-Edge infrastructures. \fogarm proved to be able to scale up to tens of nodes and hundreds of managed services while also reducing execution times and migrations thanks to \creasoning.

However, \fogarm is only an initial prototype of a next-gen orchestrator based on \creasoning to achieve the continuous and QoS-compliant management of multi-service applications on geographically distributed Cloud-Edge networks, also capable of working in continuity with the CI/CD pipeline and infrastructure monitoring. Thus, we consider here some possible limitations to our proposal. First, the current implementation of \fogarm works by interacting with Docker, but to improve the execution times it could be interesting to substitute Docker with a more advanced tool (e.g., Kubernetes) or to work directly with a container run-time environment (e.g., \cmd{containerd}), thus excluding the interactions with intermediates. At the same time, the improvement of the low-level mechanism of \fogarm should be accompanied by the development of strategies and techniques to support the stateful migrations of services, thus enabling the persistency of the data over time and nodes. Furthermore, currently \fogarm reasons only on software, IoT devices and RAM requirements. However, \fogbrain could be extended to support more expressive policies managing a richer infrastructure model including, for example, CPU, HDD and security requirements. Additionally, in the current prototype, the user has to manually insert the service's requirements. However, a useful extension could include a process of Data Mining to automatically generate, possibly exploiting a system of code's annotations, the service requirements. Finally, it would be very interesting to design such a process also exploiting a \creasoning methodology to speed up the process of requirements extraction.

To conclude, we discuss some future research lines:

\begin{description}

\item[Developing new placement strategies] Continuous reasoning is designed to boost the performance of a given placer, reducing the size of the considered problem and re-using previously computed results as much as possible. Following this principle, \fogbrain could support several different placement strategies (e.g., genetic algorithms), possibly providing either a logic programming implementation of the desired approach or a logic interface to the implementation of the placer. Furthermore, the application and infrastructure models can be enriched by considering other QoS requirements/capabilities (e.g., security properties, energy consumption). Additionally, a stateful migration mechanism could be studied to better support the application orchestration.
\item[Extending run-time decisions] Our methodologies could be extended by considering other management decisions (e.g., application scaling, service adaption), possibly including explanations on \textit{why} a certain management decision was (not) taken. This would enrich the capabilities of our orchestrator, enabling both more sophisticated application management and improving visibility into the decision-making process.
\item[Workload stress test] A further assessment of \fogarm could involve studying its behaviour under stressful conditions through increasing workload on the managed services, thus even overloading both the node and links on the infrastructure through experimenting with the actual flow of data and users' interaction even in difficult condition with service crash or data loss.
\end{description}

\section*{Acknowledgements}

Thanks are due to the GARR Consortium for allowing us to experiment with the GARR infrastructure, to the staff of the GARR Cloud Support and especially to Dr Alberto Colla for their availability and support in using GARR Cloud resources. 

\bibliographystyle{splncs04}
\bibliography{biblio}

\end{document}